\documentclass[aps,prl,showpacs,twocolumn,reprint,floats,epsfig,pdflatex,superscriptaddress]{revtex4-1}
\usepackage{tabularx}
\usepackage{graphicx}
\usepackage{caption}
\usepackage{subcaption}

\begin{document}

\title{An exacting transition probability measurement - a direct test of atomic many-body theories}

\author{T.~Dutta}
\author{D.~De~Munshi} 
\author{D.~Yum}
\author{R.~Rebhi} 
\affiliation{Centre for Quantum Technologies, National
University Singapore, Singapore 117543}
\author{M.~Mukherjee}
\affiliation{Centre for Quantum Technologies, National University Singapore, Singapore 117543}
\affiliation{Department of Physics, National University Singapore, Singapore 117551. }
\affiliation{MajuLab, CNRS-UNS-NUS-NTU International Joint Research Unit, UMI 3654, Singapore}

\date{\today}

%



\begin{abstract}
A new protocol for measuring the branching fraction of hydrogenic atoms with only statistically limited uncertainty is proposed and demonstrated for the decay of the P$_{3/2}$ level of the barium ion, with precision below $0.5\%$. Heavy hydrogenic atoms like the barium ion are test beds for fundamental physics such as atomic parity violation and they also hold the key to understanding nucleo-synthesis in stars. To draw definitive conclusion about possible physics beyond the standard model by measuring atomic parity violation in the barium ion it is necessary to measure the dipole transition probabilities of low-lying excited states with precision better than $1\%$. Furthermore, enhancing our understanding of the \textit{barium puzzle} in barium stars requires branching fraction data for proper modelling of nucleo-synthesis. Our measurements are the first to provide a direct test of quantum many-body calculations on the barium ion with precision below one percent and more importantly with no known systematic uncertainties. The unique measurement protocol proposed here can be easily extended to any decay with more than two channels and hence paves the way for measuring the branching fractions of other hydrogenic atoms with no significant systematic uncertainties.       
\end{abstract}

\flushbottom
\maketitle
%
%
\thispagestyle{empty}
%

\section*{Introduction}

Atomic and molecular structure are the fundamental building blocks for developing quantum technology, precision measurements, metrology, astrophysics and material chemistry. In particular, the lifetimes of excited states and their branching fractions upon decay to lower energy levels form the basis for our understanding of the electronic structure of atoms and molecules. In the case of hydrogen, with a single valence electron the electronic structure is exactly solvable. However, for atoms beyond hydrogen in the periodic table, the electronic structure has no exact solution due to the many body nature of the problem. Nevertheless, well developed theoretical many-body techniques make it possible to calculate electronic structure with precision below one percent for heavy atoms with an iso-electronic structure to hydrogen, {\it e.g} cesium~\cite{Dzu89, Woo97}. These types of atoms and their molecules are of particular interest in studying physics beyond the usual realm of atomic and molecular physics. The presence of a large number of nucleons in these heavy atoms influences the atomic energy levels via the electro-weak interaction~\cite{Bou97}. The singly charged barium ion is one such system where significant violation of atomic parity symmetry due to the weak interaction has been predicted but is yet to be observed~\cite{For93}. However, knowledge of the low-lying electronic states with precision below one percent is a stringent requirement but up untill now only have been known to about $5\%$ precision mainly limited by the systematic uncertainties~\cite{Kur08}. In addition, this particular ion also has astrophysical importance due to its excessive abundance in certain main sequence stars which is popularly known as the \textit{barium puzzle}~\cite{Fau09,Red15}. While in case of astrophysics precision is not a major criteria, however certain spectral lines and their relative intensities form the backbone of stellar element formation calculations~\cite{Red15}. \\
 
Methods of measuring the branching fraction of an excited atomic state have evolved over time towards better precision and lower systematic uncertainties. The earliest method is based on a ratio measurement in two branches, obtained directly from fluorescence spectroscopy~\cite{Gal67}. Particularly for the barium ion these results could not resolve the two fine-structures of the meta-stable D-state and hence two out of the three branches are only approximately known. Subsequent improvements resulted in an overall precision of about $10\%$. Kastberg~\textit{et al.}~\cite{Kas93} developed a method of measuring the branching fraction from an optical nutation experiment which provided marginal improvement over the existing results, however this method is limited by systematic constraints such as optical nutation which depends on magnetic field, laser intensity at the ion position \textit{etc.}. As ion trap technology has progressed, it has allowed single or multiple ions to be confined in a nearly perturbation free environment over long times thereby eliminating fly-by time related systematics errors. Additionally, short pulsed lasers have allowed the development of new techniques for branching fraction measurements~\cite{Kur08} with systematic uncertainties below $10\%$. The possibility of directly measuring the population in lower atomic states after being excited to the higher state with nearly $100\%$ efficiency on a single ion has allowed measurement of the branching fraction to below $1\%$ in Ca$^+$~\cite{Ger08}. This technique in principle can be used for barium where the required precision is below $1\%$. However, it requires an extra laser and the precision is limited by the ability to transfer population to the excited state in a serise of short laser pulses deterministically. Furthermore, this technique require a single ion and hence gathering statistics is time consuming. Ramm \textit{et al.} introduced a technique based on photon counting which is more robust in comparison to previous methods, and importantly, it is only statistically dominated~\cite{Ram13}. Thus arbitrary precion can be achieved by accuraing longer statistics. This resulted in first ever measurement of Ba$^+$ branching fractions below one percent~\cite{Mun15} for the P$_{1/2}$ state, where there are only two decay channels. However, direct implementation of the same technique to measure the branching fraction from an excited state with total angular momentum $j=3/2$ will be dominated by the systematics due to angular dependence of the observed photons on input laser polarization and magnetic field direction~\cite{Pru14,And73}. In this article, we propose and implement a new technique which can provide the robustness of photon counting as well as low uncertainties dominated only by measurement statistics. \\

\section*{Results}
\label{sec:meth}
\subsection*{Proposed protocol}

\begin{figure}
\includegraphics[width=0.8\linewidth]{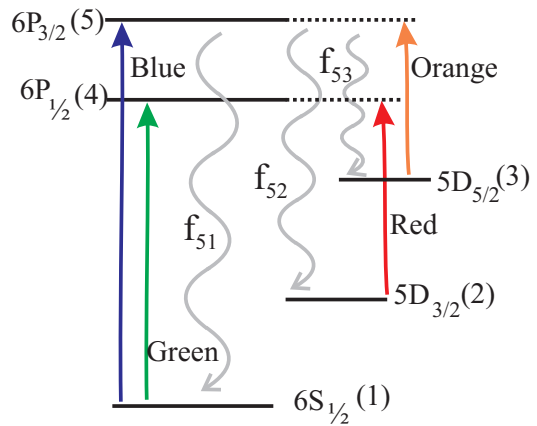}
\caption{(Color online) Relevant energy levels of a hydrogenic atom: The excitation lasers are generically referred to as green, blue, red and orange while the relevant levels are denoted as $1..5$ as shown in bracket. In particular the decay channels from the P$_{3/2}$ level~(5) are of importance.}
\label{Fig1}
\end{figure}

Let us consider the relevant energy levels of hydrogenic atoms as shown in figure~\ref{Fig1}. The P$_{3/2}$ state decays into three branches D$_{5/2}$, D$_{3/2}$ and S$_{1/2}$ which will henceforth be referred to as $3$, $2$ and $1$ respectively. The upper two P-states with $J=1/2$ and $J=3/2$ are referred to as $4$ and $5$ respectively. In this article, the lasers that excite four possible dipole transitions shown in fig.~\ref{Fig1} are generically referred to as green, blue, red and orange. In particular for the barium ion, these transitions are at wavelengths $493$, $455$, $650$ and $614~$nm respectively. The branching fractions $f$ of the P$_{3/2}$ state as it decays into the lower atomic states $1$, $2$ and $3$ are proportional to these states' decay probabilities which are in turn proportional to the square of their respective transitional dipole matrix elements. These branching fractions are constrained as  

\begin{equation}
f_{51}+f_{52}+f_{53}=1.
\label{eq1}
\end{equation}

If one of the three decay channels is blocked, say for example channel $1-5$ by continuous blue laser excitation (denoted by superscript $b$), the probability ratio r$_{32}$ for the atom to be found in the other two levels, $3$ and $2$ after decaying from $5$ is expressed as 

\begin{equation}
f_53/f_52 = f^b_{53}/f^b_{52} = r_{32},
\label{eq2}
\end{equation}

where $f^b_{5i=2,3}$ is the $i$th state decay fraction under blue channel blocked condition. Since $f^b_{52}+f^b_{53}=1$, the ratio can be expressed as

\begin{equation}
r_{32} = (1-f^b_{52})/f^b_{52}.
\end{equation}   

Thus one-colour-photon measurement of $f^b_{52}$ provides complete information about the ratio $r_{32}$. In a similar approach but starting with the entire population in state $3$, by blocking orange channel $3-5$, it is possible to measure $r_{12} = f_{51}/f_{52} = f^o_{51}/f^o_{52}$ by measuring the population in the state $2$ alone. 
Therefore it is straight forward to show that 

\begin{equation}
f_{52} = 1/(r_{12}+r_{32}+1).
\label{eq3}
\end{equation} 

The main goal of this protocol is to measure the ratios $r_{12}$ and $r_{32}$ with low systematic uncertainties. Both ratios depend on the probability of finding the atom in state $2$, and the population in this state can be measured without any significant systematic error as it is the lowest excited state~\cite{Mun15}. This is done by repumping the population from $2$ to $4$ while counting the number of spontaneously emitted photons between $4-1$ (green photons) per cycle, and measuring the total efficiency of this photon counting. The efficiency can be measured by counting the green photons when the atom is re-pumped from state $2$ to $4$ after having completely transferred to state $2$ starting from the ground state $1$ by application of the green pump laser. In this case the theoretical number of green photons emitted per cycle is exactly $1$, hence the efficiency $\epsilon$ is determined by counting the total number of green photons $N^m_{green}$ emitted in $c$ cycles where the subscript $green$ refers to the pumping laser. Thus,

\begin{equation}
\epsilon = N^m_{green}/c.
\label{eq4}
\end{equation}    

As an example, the population transfer sequence for the measurement of $r_{32}$ starts with one or a few ions in their ground electronic state $1$. A complete transfer to the D-states ($2$ and $3$) is done by applying the blue laser driving the $1\rightarrow 5$ transition (pumping step). The red laser is then applied to completely transfer any population from state $2$ to $1$ via state $4$ (re-pumping) and the number of emitted green photons ($4\rightarrow 1$) is counted. This sequence repeated for $c$ cycles provides $N_{blue}$ green photon counts (where the subscript again refers to the pumping laser). Therefore, 

\begin{equation}
r_{32} = \frac{\Big(1-\frac{N^m_{blue}/c}{\epsilon}\Big)}{\frac{N^m_{blue}/c}{\epsilon}} = \frac{\Big(1-\frac{N^m_{blue}}{N^m_{green}}\Big)}{\frac{N^m_{blue}}{N^m_{green}}}, 
\label{eq5}
\end{equation}    
  
where the efficiency is cancelled out and the ratio is independent of any external parameters used in the experiment. Similarly, starting from state $5$ and applying the orange pump laser for complete population transfer to states $1$ and $2$, the ratio $r_{12}$ can be expressed as

\begin{equation}
r_{12} = \frac{\Big(1-\frac{N^m_{or}}{N^m_{green}}\Big)}{\frac{N^m_{or}}{N^m_{green}}}, 
\label{eq6}
\end{equation} 

where, $N^m_{or}$ refers to the number of green photons emitted in $c$ cycles while blocking the $5\rightarrow3$ decay channel by shining an orange laser light. \\


\begin{figure}
\includegraphics[width=0.9\linewidth]{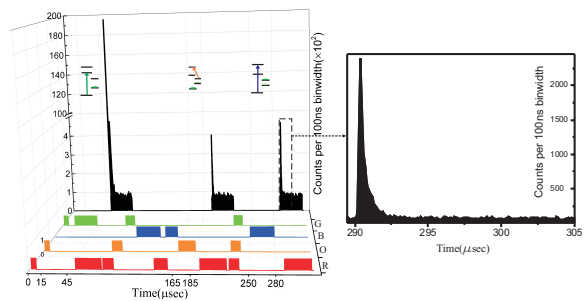}
\caption{(Color online) Experimental result: (left) The laser pulse time sequence for population shutteling between the states are ploted along with the observed green photon counts during the red laser re-pumping stage. The time sequence for red~(R), orange~(O), blue~(B) and green~(G) are shown sequentially. The total number of green photons emitted during the red re-pumping stage are experimentally measured in time bins of $100$~ns as shown on the back side of the time sequence plot. (right) A zoomed in plot for one such decay shows that after re-pumping the emission level of the green photons reach background level.} \label{Fig3}
\end{figure}


Possible systematic effects that may influence the measured branching fractions in any photon counting protocol come from the dead time of the detector, usually a photomultiplier tube. The dead time introduces a systematically lower count than the actual count if the photomultiplier tube is operated in a non-linear response regime where more than one photon arrives at the detector within its dead time. However, in the scheme presented here, this is not possible due to the technique of projective measurement which means that only one green photon can arrive at the detector per cycle, once it is projected by the red laser unless there are chance coincidences due to background photons of same colour. The number of emitted photons from an oscillating dipole depends on the quantization axis (magnetic field direction in our experiment), polarization of the exciting laser field and the total angular momentum of the excited state. Since the proposed protocol measures the photon counts from the P$_{1/2}$ level only, the total count is insensitive to input laser polarization, or in other words the Hanle effect does not introduce any systematic error, provided the detection setup is circular polarization insensitive which is usually the case. Experimentally, the influence on photon counts due to the input polarization can be quantified. As the scheme relies on photon counting rather than state detection, population transfer is always complete for any given transition. The experiment is not affected by the number of ions provided the ion number remains the same for each measurement cycle, however it is essential to keep the ions in a crystalline state so that the probed ions do not drift out of the field of view of the photon collection optics. In the following section, systematic uncertainties related to our experiment are further elaborated. Most importantly, this scheme is scalable to any decay with more than two channels, since branching fractions can be measured with reference to the lowest excited state where the efficiency is well calibrated. Therefore it can be easily applied to other heavy atoms like Yb$^+$, Sc$^{++}$, Ra$^+$ and Hg$^+$.

\subsection*{Measurement precision and error budget}

The measurement technique when applied to the barium ion P$_{3/2}$ state is based on counting $493$~nm (green) photons that reach the detector within an interval of time in which the red laser is on. An interference filter with narrow bandwidth around $493~$nm placed in front of the PMT results in background free measurement of the photon counts. The characteristic feature of this count as a function of time is an exponential decay as shown in figure~\ref{Fig3}. The total area under each curve provides the total photon number corresponding to N$_{green}$, N$_{or}$ and N$_{blue}$ respectively as in eq.(\ref{eq5}) and eq.(\ref{eq6}). In order to extract the ratios $r_{12}$ and $r_{32}$, background counts (mostly dark counts) taken within the same time interval as the measurement have been subtracted. The background counts are taken in each cycle just after each measurement so as to ensure similar ambient condition. The result of the three branching fractions measurement are summarized in table~\ref{tab1}. 

\begin{table*}
  \caption{Error budget for the P$_{3/2}$ branching fraction measurement. The uncertainties mentioned along with the fraction are the total uncertainties (systematic and statistical). A break down of the total uncertainties from different sources are shown above the final values. The photomultiplier tube used is from Hamamatsu model no. H7421-40 for which we measured the dead time and its uncertainty.}
  \begin{tabularx}{1\textwidth}{lXXXXXX}
    \hline\hline
    
    {\bf parameter} & \multicolumn{3}{c}{\bf Shift} & \multicolumn{3}{c}{\bf Uncertainty} \\
    				&  $5\rightarrow1$ & $5\rightarrow2$ & $5\rightarrow3$ &  $5\rightarrow1$ & $5\rightarrow2$ & $5\rightarrow3$ \\
    \hline
    Statistical			& & & & $0.00169$ & $0.000175$ &  $0.00154$ \\
    Detector dead time $54.5(0.5)~ns$\quad  &$1.09\times10^{-4}$ & $1.35\times10^{-4}$& $2.06\times10^{-5}$ &$2.39\times 10^{-5}$ & $6.24\times 10^{-7}$ & $8.91\times10^{-6}$ \\
	Finite measurement time & & & &$7\times 10^{-7}$ & $3\times 10^{-8}$ & $8\times10^{-7}$ \\
\hline
{\bf Transitions} & \multicolumn{2}{c}{\bf $5\rightarrow 1$} & \multicolumn{2}{c}{\bf $5\rightarrow 2$} & \multicolumn{2}{c}{\bf $5\rightarrow 3$}\\
\hline
	Final branching fraction	& \multicolumn{2}{c}{$0.7380\pm0.0017$} & \multicolumn{2}{c}{$0.02862\pm0.00017$} & \multicolumn{2}{c}{$0.2333\pm0.0015$}\\
	\hline\hline	 
  \end{tabularx}
  \label{tab1}
\end{table*}               
      
In any precision measurement the most important experimental challenge is to identify and measure all possible systematic uncertainties associated with the experiment. In the following, all possible systematic effects associated with our experiment are discussed in detail:


\begin{figure*}
\centering
\begin{subfigure}{0.4\linewidth}

\includegraphics[width=0.9\linewidth]{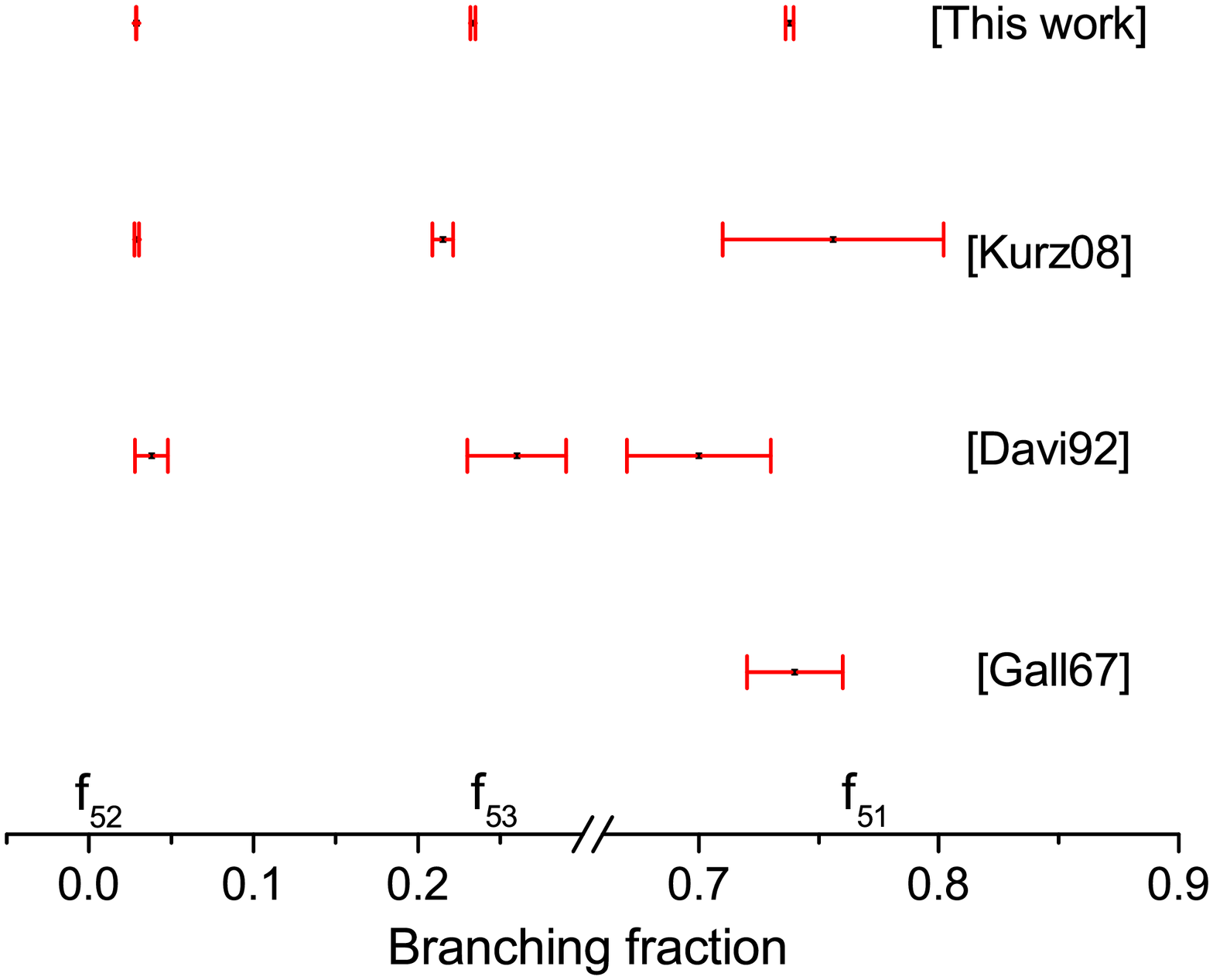}
\end{subfigure}
\begin{subfigure}{0.4\linewidth}
\centering
\includegraphics[width=1\linewidth]{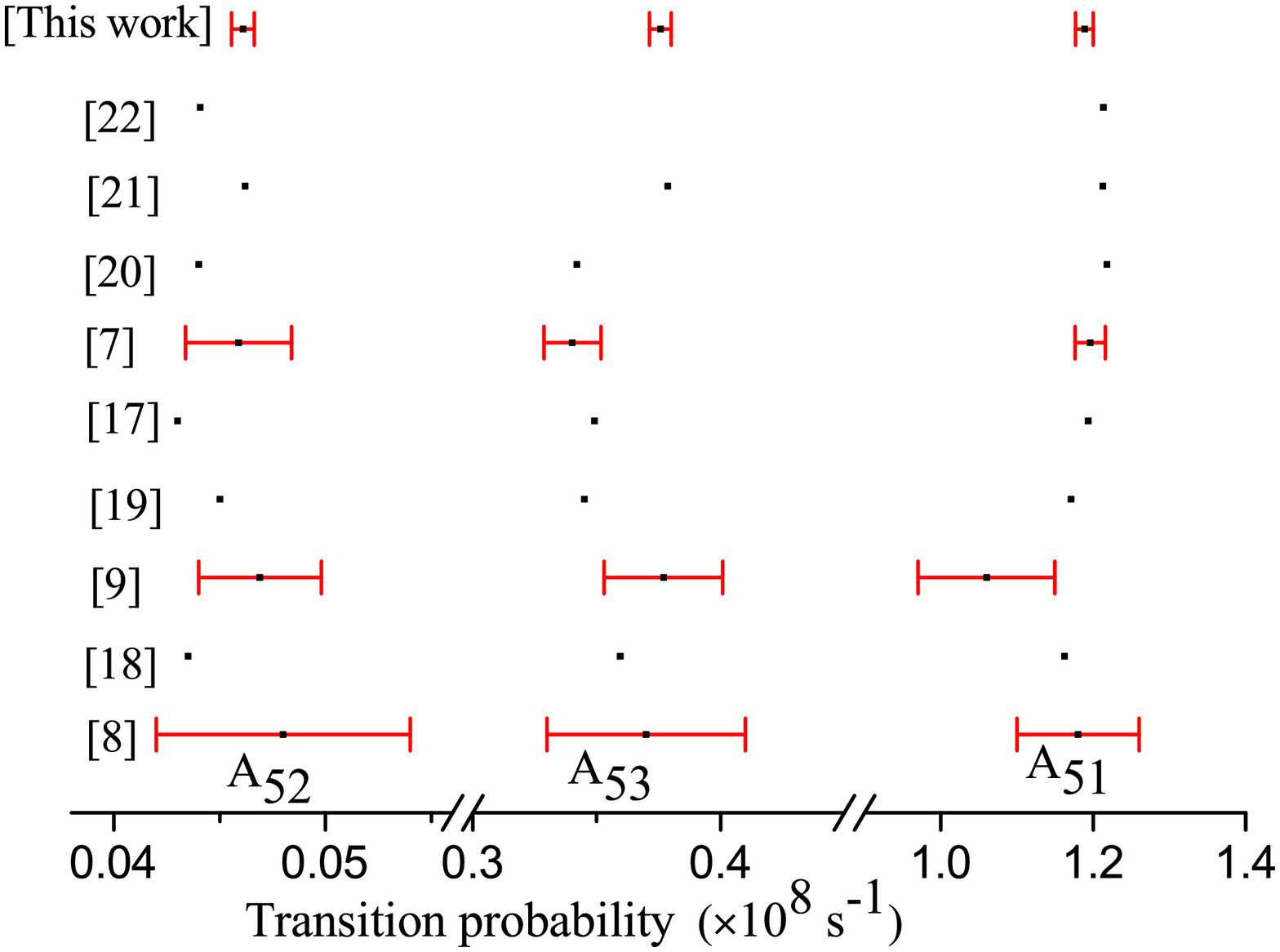}
\end{subfigure}
\caption{Comparison of measured branching fraction for the P$_{3/2}$ state decay of Ba$^+$ (left): Different experimentally measured values of the branching fraction and their uncertainties are plotted. For ref.~\cite{Gal67} the fraction decaying to the D-state were not measured. (right) Comparative plot of transition probabilities measured or calculated by different groups over many decades are plotted in a chronological manner from bottom to top. In case of the theoretically calculated values, uncertainties are not mentioned. The plot shows how precision experiments enhanced how theoretical understanding of the quantum many-body calculations.}
 \label{Fig4}
\end{figure*}


\begin{itemize}

\item \textit{PMT dead time:} In a single photon counting regime, once a PMT registers a single photon it remains dead for a certain time before it can register a second photon. Therefore in our experiment, if the count rate is high such that successive photons arrive at the detector within the dead time, one must account for the loss in photon counts. It is straight forward to calculate such a loss and the resulting shift of the branching fraction, as well as the uncertainty, which depends on the uncertainty of the measured dead time~\cite{Mun15}. However, the technique proposed here is devoid of such a shift and associated uncertainty since there is at the most one green photon emitted per cycle of the re-pumping red laser. Relying on projective measurement, this technique ensures a single green photon emission before the ion is projected to the ground state and hence decoupled from the exciting red laser~\cite{Ram13}. Table~\ref{tab1} provides an uncertainty related to the dead time which is an estimate derived from the chance coincidences of having a background photon and photon emitted by an atom arriving at the detector within the detector dead time. We have conservatively considered the shift as an uncertainty itself. As the photon counting is done with no background laser at the emitted wavelength, this uncertainty is negligibly low. 

\item \textit{Finite measurement time:} In the measurement scheme employed here, the total number of registered photons are calculated from the observed decay curve after subtracting the background. Since a decay curve ideally extends to infinity and the measurements are taken within a finite time, one needs to account for possible loss of counts. This has been estimated by fitting the decay curves with the proper exponential and conservatively assuming the uncertainty to be the same as the shifts. As shown in figure~\ref{Fig3}, the applied red laser pulse is about $15$ times longer than the characteristic decay time, therefore this contribution is well below the statistical uncertainty as shown in table~\ref{tab1}.

\item \textit{Polarization dependence:} In case the ions are excited by circularly polarized light and the detection system is more sensitive to one type of circular polarization of the emitted photons than the other, the observed count rate will then be a function of the magnetic field provided the field has a component perpendicular to the direction of propagation of light. This is known as the Hanle effect. In the case of a measurement scheme as proposed in~\cite{Ram13} but applied to the decay of $J=3/2$ state, this effect would be the largest contributor to the systematic uncertainties~\cite{Pru14}. However, the scheme proposed here is devoid of it as the measurement is done using only P$_{1/2}$ state decay. Since our detection system is identical to that used in~\cite{Mun15}, the measured effect of polarization asymmetry in our detection setup is below the statistical uncertainty of the measurement. We further confirmed this, by measuring the branching ratios for different input polarizations, for equal statistics.

\item \textit{Collision and radiative decay:} Once a barium ion is excited by the blue light, it decays to two meta-stable D-states and the ground state. The population of the meta-stable states must be measured before the ion decays to the ground state otherwise the measured branching fraction of the meta-stable states will be systematically low. The decay of meta-stable states can be radiative or the states can quench due to collisional energy transfer. These two issues have been a major challenge in the previously performes experiments~\cite{Kur08}. However, in our case since the total measurement cycle period is only $320~\mu$s and the meta-state states have lifetimes of $30$ and $80$~seconds, the meta-stable states are populated at most for $20-30~\mu$s during one cycle. The skewing of results due to radiative decay is therefore estimated to be well below statistical uncertainties. Similarly, the collisional decay rate at the experimental working vacuum of $10^{-10}$~mbar, is estimated to be negligible. Considering a collisional quenching rate of $84\pm11\times 10^3$ s$^{-1}$Pa$^{-1}$ due to H$_2$ which is the dominant background gas in our experiment~\cite{Mad90}, leads to a rate of $42$~mHz with a conservative pressure estimate for the center of the trap. Therefore, this effect is negligible within a time window of $30\mu$s. For elements like Ca$^+$ with relatively short lived meta-stable states, it is important to measure this decay as it may add to the systematics. However, the total cycle period can easily be shortened further by faster population transfer with higher laser power. We further verified experimentally that both the D-states do not decay within our experimental shelving times by probing the shelved states after different waiting times.

\item \textit{Collisional mixing:} Collision with background gas molecules not only causes decay of the meta-stable states but also leads to mixing of the fine-structure levels~\cite{Kno95}. This may lead to population transfer between the D-states. The collisional mixing can be modelled by Landau-Zener theory which leads to the result that the rates depend inversely on the energy difference between the fine-structure levels. No experimental data is available on this mixing rate for barium, however it can be estimated from measured values for the calcium ion~\cite{Kno95}. Since both calcium and barium are hydrogenic atoms and the D-state mixing due to collision is dependent on electronic structure, particularly on the energy difference between the fine-structure levels (the barium ion has $11$ times larger separation between the fine-structures than the calcium ion), the rate of collisional mixing in barium can be extrapolated from calcium rates. Following ref.~\cite{Kno95} this mixing rate is calculated to be $2\times10^{-7}$~Hz which means that D-state mixing is negligible in our case. Note that D-state mixing is not negligible for the Ca$^+$ ion, where it may lead to a significant systematic error provided the operational vacuum is worse than $10^{-7}$~mbar. In a similar way as mentioned in the previous bullet point, we also verified that the states are not mixing within our experimental cycle time.

\item \textit{Laser extinction:} During the off state of any laser, we achieve better than $70$~db suppression as compared to the on state owing to the double pass configuration. Therefore, experimentally we have not observed any population transfer during off state of the lasers.          

\end{itemize} 

After considering all possible error sources, the final value of the branching fractions and the total uncertainties are tabulated in table~\ref{tab1}. As a comparison, previously measured values of the branching fractions are shown in figure~\ref{Fig4}. Our results match well with previously reported values for $5-1$ and $5-2$ transitions but with significantly better precision by a factor of $10$ or more. In the case of the transition $5-3$, we obtain a branching fraction which is higher than the previously reported best value~\cite{Kur08}, but our result falls within the errorbars of earlier result~\cite{Gal67,Dav92} which has larger uncertainty. The uncertainty of our measurement is $6$ times lower than the previous ion trap measurement~\cite{Kur08} which was limited by the decay of the meta-stable states. Since our measured transition probability is higher compared to the previous measurement, we believe that the small discrepancy may come from an under-estimate of the upper state decay in the previous experiment, which is known to contribute to the systematic uncertainty of that measurement protocol. 
%
%

\subsection*{Transition probabilities and matrix elements}

The dipole allowed transition probability between two states depends on the upper state life-time and the branching fraction to the lower state given by:

\begin{equation}
A_{i\leftarrow k} = f_{i\leftarrow k}/\tau,
\label{eq7}
\end{equation}  

where $A$, $f$ and $\tau$ denotes the transition probability expressed as $s^{-1}$, branching fraction and upper state ($k$) life time in $ns$. Moreover, the transition probability is related to the reduced dipole matrix element between the involved states, denoted as $R$, by~\cite{Gee02}:

\begin{equation}
A_{i\leftarrow f} = \frac{2.0261\times 10^{18}}{g_k\lambda^3}|R|^2,
\label{eq8}
\end{equation}

where $g_k$ and $\lambda$ stand for the multiplicity of the upper state $k$ and the vacuum wavelength expressed in $\AA$. 
The result of a many-body theory calculation from first principles yields the eigen states of the atom or ion. Therefore, theorist usually provide the value of $R$ or $A$ while experiments provide the measured value of $f$. Using the measured $f$ and the best known value of $\tau$~\cite{And73} to be $6.21(06)~$ns, we calculated the value of $A$. Note that the uncertainty on the calculated $A$ depends on the uncertainties of both $f$ and $\tau$. From eq.~(\ref{eq8}), it is straight forward to calculate $R$, for which the best known value of the wavelength, recommended by the Nation Institute of Standard and Technology (NIST), has been used. A comparative study of the transition probabilities $A$ obtained from previous experiments and existing theory calculations, shown in figure~\ref{Fig4}, indicates how experiments have evolved over time to catch up with the rapid development of many-body techniques employed to calculate complex systems like the barium ion with as many as $111$ electrons and protons. While theoretical advancement has been powered by new approaches like coupled-cluster theory and the availability of super-computers~\cite{Gue91, Dzu01, Gee02, Sah07, Saf10, Dut14}, experimental advancement has been driven by improvements in ion trap technology~\cite{Kur08,Ger08} and refined methodologies of quantum state manipulations~\cite{Ram13}. Our results for all three branches are only limited by the uncertainty of the upper-state lifetime, which is not a part of our present experiment. However, in spite of this uncertainty, our results allow for the first direct measurement of branching fractions for the barium ion with precision below one percent which is the benchmark for possible atomic parity violation experiments. It is further to be noted that our measurement uncertainties are only limited by statistics and hence can always be improved by gathering more statistics. As seen in figure~\ref{Fig4}, the results presented here match well with previous experiments and are within the uncertainties for the transitions $5\rightarrow 1$ and $5\rightarrow 2$, with improved uncertainty by a factor of about $5$. Note that the calculated uncertainty in ref.~\cite{Kur08} of $A$ for the transition $5\rightarrow 1$ is $1.6\%$ which is about $4$ times lower than the measured uncertainty of $f$ in that experiment. The transition probability $A_{3\leftarrow 5}$ however, measured to be higher as compared to ref.\cite{Kur08}, while the result of our measurement are within the uncertainties of other previous measurements~\cite{Kas93}. 

\begin{table*}
  \caption{Comparison of measured and calculated transition probabilities for the decay P$_{3/2}$ state of barium ion. The transition probabilities for ref.~\cite{Saf10,Dut14} are derived using NIST recommended wavelength values.}
  \begin{tabularx}{0.7\textwidth}{lXXX}
    \hline\hline
    {\bf transition} & {\bf transition probability A$_{i\leftarrow k}$} & {\bf reduced matrix element R$_{ik}$} & {\bf Reference} \\
					&	($\times 10^8$~s$^{-1}$) & (a.u) & \\
    \hline
    $1\leftarrow 5$			& $1.188\pm0.012$ & $4.709\pm0.033$ & This work \\
							& $1.2134$		&	$4.7586$ & \cite{Dut14} \\
							& $1.2125$		& 	$4.7569$ & \cite{Saf10} \\
							& $1.218$		&	$4.72$ 	& \cite{Sah07} \\
							& $1.196\pm0.02$ & $4.72\pm0.04$ & \cite{Kur08} \\
							& $1.1937$		& $4.6982$ & \cite{Gee02} \\
							& $1.171$		&	$4.6746$ & \cite{Dzu01} \\
							& $1.06\pm0.09$ & 			& \cite{Kas93} \\
							& $1.18\pm0.08$ &			& \cite{Gal67} \\
\hline
    $2\leftarrow 5$			& $0.04610\pm0.00053$ & $1.3517\pm0.0077$ & This work \\
							& $0.04408$		&	$1.3217$ & \cite{Dut14} \\
							& $0.04620$		& 	$1.3532$ & \cite{Saf10} \\
							& $0.044$		&	$1.34$ 	& \cite{Sah07} \\
							& $0.04589\pm0.0025$ & $1.349\pm0.036$ & \cite{Kur08} \\
							& $0.043$		& $1.2836$ & \cite{Gee02} \\
							& $0.045$		&	$1.3346$ & \cite{Dzu01} \\
							& $0.0469\pm0.0029$ & 			& \cite{Kas93} \\
							& $0.048\pm0.006$ &			& \cite{Gal67} \\
\hline							
    $3\leftarrow5$			& $0.3758\pm0.0044$ & $4.1475\pm0.0034$ & This work \\
							& $-$		&	$-$ & \cite{Dut14} \\
							& $0.3786$		& 	$4.1631$ & \cite{Saf10} \\
							& $0.342$		&	$4.02$ 	& \cite{Sah07} \\
							& $0.3402\pm0.0115$ & $3.945\pm0.066$ & \cite{Kur08} \\
							& $0.349$		& $3.9876$ & \cite{Gee02} \\
							& $0.345$		&	$4.1186$ & \cite{Dzu01} \\
							& $0.377\pm0.024$ & 			& \cite{Kas93} \\
							& $0.37\pm0.04$ &			& \cite{Gal67} \\						
	\hline\hline	 
  \end{tabularx}
  \label{tab2}
\end{table*}

If we compare the present results with those of theoretical calculations it is now possible to clearly distinguish between different theoretical results with a precision of $1\%$. Theoretical calculations commonly employ relativistic coupled cluster theory to all orders in perturbation, but with different orders in excitations and in some cases with added interactions like Breit corrections. The calculated values of the transition probabilities also reply on the vacuum wavelength indicated in eq.~(\ref{eq8}), and some of the theory values use calculated energy eigen values for the wavelength which are about $2-3\%$ different from the NIST recommended values~\cite{Cur04}. Therefore, it is now essential to understand the choice of initial wavefunction so as to improve the calculations such that the calculated eigenvalues and eigenfunctions can meet the $1\%$ requirement of a parity violation experiment on barium ion.

\section{Discussion}

An improved protocol to measure the branching fractions of a hydrogenic atom with more than two decay channels has been proposed. It is shown that this protocol has no significant systematic uncertainty and hence the final precision of the measurement is only limited by statistics. Using the new protocol, measurements performed on the barium ion P$_{3/2}$ state have for the first time allowed the branching fractions to be measured with precision below one percent. The resulting calculated values of the transition probabilities are now only limited by poor knowledge of the P$_{3/2}$ lifetime. Therefore, for the barium ion these values clearly sets the test bed for high precision many-body atomic structure calculations to be performed with precision below one percent.

A comparison of calculations performed by different theory groups of the dipole matrix elements, with our experimental results, shows discrepancies. These may originate from inaccurate starting wavefunctions which systematically provide energy eigenvalues of hydrogenic atom that differ from experimental values by about $2-3\%$. Another possibility may be due to underestimated terms in the Hamiltonian. In either case, the availability of measured values with precision better than $1\%$ will help better understand many-body quantum physics. Moreover, these measurements along with the measurements performed in~\cite{Mun15} are nearly sufficient towards the total dipole matrix element contribution of the atomic parity violation Hamiltonian~\cite{Dzu01}. Unlike in the case of Cs, the atomic parity violation contribution due to mixing of the higher P-states in Ba$^+$ are not cancelled out by sum-over-states calculations. Therefore our measurements will help pinning down the accuracies of the theoretical results on the violation of atomic parity in the barium ion~\cite{Sah07}.       
 
\section*{Method}


\begin{figure}[b]
\centering
\includegraphics[width=0.9\linewidth]{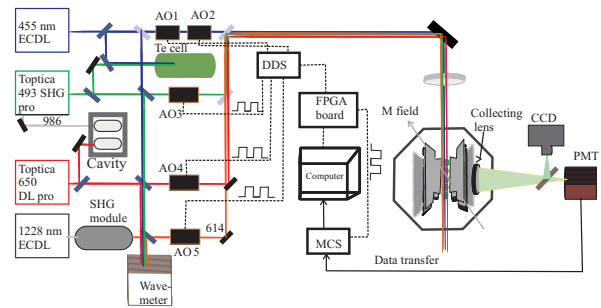}
\caption{(Color online) Schematic diagram of the experimental setup. Four different lasers along with their locking schemes are shown by solid coloured lines while the electronic control paths are shown by dashed lines.} \label{Fig2}
\end{figure}


The experiment schematically shown in figure~\ref{Fig2} consists of a linear Paul trap which is operated at a radial frequency of about $1$~MHz while the axial frequency is below one $1$~kHz depending on the number of crystallized ions. This experiment involves a total four lasers for quantum state manipulation and one more for photo-ionization. All the lasers are diode lasers in external cavity configuration to ensure narrow linewidth and stability. As shown in figure~\ref{Fig2} the green (493 nm) and the blue (455 nm) lasers are frequency locked to molecular Te$_2$ spectra separately while the red (650 nm) laser is phase locked to a reference cavity as is the master (986 nm) laser of the green laser. The setup is identical to the one in ref.~\cite{Mun15} with the exception of two additional diode lasers: one for driving the S$_{1/2}$-P$_{3/2}$ transition at $455$~nm (blue) and the other one to drive the D$_{5/2}$-P$_{3/2}$ transition at $614$~nm (orange). The blue laser is made from a commercially available $450$~nm diode which is frequency locked to a Te$_2$ transition which has been observed for the first time and reported in ref.~\cite{Dut16}, similar to what has been done for the $493$~nm laser. The $614$~nm laser is a frequency doubled laser produced by an optical-fibre based periodically polled crystal in single pass configuration from a master laser at $1228$~nm. The master laser is an external cavity diode laser (ECDL) using a commercial gainchip. Both the amplitude and frequency of all four lasers are individually controlled by separate acousto-optic~(AO) modulators which in turn are controlled by direct digital synthesizers~(DDS). Amplitude modulation is performed by a Field-Programmable Gate Array (FPGA) controlled by a computer code written in python. The laser lights are mixed using a combination of dichroic mirrors and polarizing beam splitters. The combined beam passes through the linear trap at an angle with respect to the axial electrodes. During the course of this experiment, ion numbers ranging from a few to about $10$ were cooled down to the Doppler limit temperature using the green and red lasers simultaneously. The ions are detected by fluorescence emitted at $493$~nm perpendicular to the laser beam path with an optical collection solid angle of about $4\%$.  A narrowband filter at $493$~nm($\pm3$~nm) placed in the detection path ensures collection of only $493$~nm photons. The photons are detected by a Hamamatzu photomultiplier tube (PMT) with a quantum efficiency of about $40\%$ for green photons, and counted by time binning their arrival time using a multi-channel scalar (MCS) from ortec. A typical experimental cycle shown in figure~\ref{Fig3} consists of a short cooling, green pumping, red measurement, cooling, population transfer to D$_{5/2}$ using blue and red, orange re-pumping, red measurement, cooling, blue pumping and finally the red laser measurement. The total cycle time is about $320~\mu$s while there are three cooling steps within an individual cycle period to ensure non-melting of the ion chain crystal. In order to obtain the required statistics, each experiment consists of $3,000,000$ cycles, and we performed $18$ such experiments with similar numbers of cycle in order to analyse systematic uncertainties.

%
%

%
%
%

%

\bibliography{sample}


\section*{Acknowledgements}

MM would like to acknowledge careful proof reading of the manuscript by Noah Van Horne. DY and MM would like to acknowledge the financial support from AcRF MoE under the project MOE2014-T2-2-119 while TD and DDM are supported under MOE/NRF, Singapore grant. 

%
%
%

%
%
%
%

\end{document}